\documentclass[12pt]{article}   
\usepackage{graphicx}
\newcommand{\half}{\frac12}

\newcommand{\beq}{\begin{eqnarray}}
\newcommand{\eeq}{\end{eqnarray}}

\def\){\right)}
\def\({\left( }
\def\]{\right] }
\def\[{\left[ }

\def\tr{{\rm\ Tr}}

\def\eps{\epsilon}

\def\be{\begin{equation}}
\def\ee{\end{equation}}
\def\bea{\begin{eqnarray}}
\def\eea{\end{eqnarray}}

\newcommand{\rem}[1]{}
\def\tr{{\rm tr}}
\def\half{{1\over 2}}

\newcommand{\abs}[1]{{\left|#1\right|}}

\def\ltap{\ \raise.3ex\hbox{$<$\kern-.75em\lower1ex\hbox{$\sim$}}\ }
\def\gtap{\ \raise.3ex\hbox{$>$\kern-.75em\lower1ex\hbox{$\sim$}}\ }

\newcommand{\ba}{\begin{eqnarray}}
\newcommand{\ea}{\end{eqnarray}}
\newcommand{\no}{\nonumber \\}

\def\bea{\begin{eqnarray}}
\def\eea{\end{eqnarray}}

\def\half{\frac{1}{2}}

\def\eps{{\epsilon}}
\def\Tr{{\rm Tr}}
\def\tr{{\rm tr}}
\begin{document}
 
\begin{titlepage}

\begin{center}
\hfill hep-th/0106093
\vspace{1cm}

{\Large\bf  Orientifold in Conifold and Quantum Deformation}

\vspace{2cm}
{\large Changhyun Ahn$^\dagger$, Soonkeon Nam$^\ddagger$, {\small and} Sang-Jin Sin$^*$}
  
\vspace{1cm}

{ \it $^\dagger$ Department of Physics, Kyungpook National University, Taegu, 702-701, Korea\\
\vskip.2cm
  $^\ddagger$Department of Physics and Basic Sciences Research Institute, \\ 
   Kyung Hee University, Seoul, 130-701, Korea\\
\vskip.2cm
  $^*$ Department of Physics, Hanyang University, 
        Seoul, 133-791, Korea }       

\end{center}
\vspace{1cm}

\begin{abstract}
We describe orientifold operation defining O3 plane in the conifold background by deriving it 
from that of O4 plane in the Type IIA brane construction by  T-duality. 
We find that both  $O3^+$ and  $O3^-$ are at the tip of the cone 
so that there is no net untwisted RR charge. 
RG analysis shows that we need two `fractional' branes for the conformal
invariance in orientifolded conifold. 
We argue that the gravity solution is the same as Klebanov and Tsyetlin since 
SUGRA  cannot distinguish the orientifolds and  D branes in this case.  
We describe the  duality cascade as well as the quantum deformation of the 
 moduli space of the field theory in the presence of the orientifold. 
 The finitely resolved conifold does not allow the orientifold, while 
deformed conifold leaves us an unresolved issue on supersymmetry.  
\end{abstract}

\vskip 2cm
\hrule
{\small \tt ahn@bh.knu.ac.kr, nam@khu.ac.kr, sjs@hepth.hanyang.ac.kr}
\end{titlepage}

\newpage
\section{Introduction}
A fruitful generalization of the duality between ${\cal N}=4$
$SU(N)$ gauge theory and type IIB strings on $AdS_5 \times S^5$
  \cite{Maldacena, GKP,WittenADS} is to consider other backgrounds
of type IIB string theory, say on $AdS_5 \times X_5$, where $X_5$
is a positively curved Einstein manifold.  We are
interested in theories with reduced supersymmetry, variety of gauge groups and matter contents. 

The simplest example realizing the ${\cal N}=1$ super symmetry is 
provided by D3 branes at the singularity of a Calabi-Yau threefold
with singularity known as conifold \cite{Candelas:1990js}. 
The conifold in the AdS/CFT context was first considered in \cite{Kehagias,KW} 
and generalized  to the case where conifold is deformed by the quantum effect of the 
fractional branes by Klebanov and Strassler  \cite{kl+st}. 
 
To construct more realistic model, gauge theory with 
$SO(N)/Sp(N)$ is necessary and this can be obtained in brane 
language by including the orientifolds  \cite{Sagnotti,Gimon:1996rq}.  
Since there are so many $Z_2$ symmetry that can act on the conifold  \cite{KW}, 
one needs care to determine which symmetry is relevant to the specific 
orbifolding operation to define the orientifold. 
One way to fix the notion of the orientifold operation is to try 
to derive it from that of the type IIA brane construction where 
things are canonically defined. 
 Under the T-duality the regular D4 branes are mapped to the regular D3 branes and 
 and the fractional D4 branes are mapped to fractional branes which can be considerred as D5 branes
 wrapping the the vanishing a 2-cycle.  
 We  use ${\rm T}$ duality using the prescription 
 given in \cite{DasM}.  

There are discussions on orientifolding the conifold based on O6 plane in type IIA picture
 \cite{Park:2000ep,oh,ohhyun,Sinha:2000ap,Naculich:2001xu}.
Here we present a discussion on
the orientifold in the conifold based on O4 brane of type IIA, because that is the one relevant to
the physics of fractional branes discussed in Ref.\cite{kl+st}. 
However it should be kept in mind that the resulting O3 branes has nature of 
 fractional brane, since in type IIA picture O4 branes are between NS 
 branes like D4 branes that is mapped to fractinal D3 branes.  
Therefore our resulting O3 branes has the character of the wrapped 
O5 branes, although our starting point in type IIA is completely different 
from the O6 brane configurations.

We will find that both  $O3^+$ and  $O3^-$ are at the tip of the cone without annihilating each
other since they are 'topologically protected' from mutual annihilation. 
But their effect is combine to give zero net untwisted RR charges.  
The non-perturbative quantum effect  deforms the tip of the cone. We will discuss how 
orientifold response to the deformation or resolution of the  conifold.
The finitely resolved conifold does not allow the orientifold, while 
deformed conifold leaves us an unresolved issue on supersymmetry.
We will also show that we need two `fractional' branes for the conformal
invariance so that the theory should have gauge group $SO(N+2) \times Sp(N)$.  
We will give several explanations for this.
We then give generalizations of the gauge theory results 
of ref. \cite{kl+st} to the case with orientifold. We 
 describe the  duality cascade and chiral symmetry breaking 
as well as the quantum deformation of the  moduli space of the field 
theory in the presence of the conifold. 
We argue that the gravity solution is the same as Klebanov and Tsyetlin since 
it cannot distinguish the orientifolds and  D branes.   

The rest of paper comes in following order.
In section 2, we give  a brief review on the relevant background.  
In section 3, we derive the 
orientifold operation from that of Type IIA picture. 
In section 4, we perform renormalization group analysis to  fix the 
that the conformally invariant
configuration as well as to compare bulk and boundary theory.
We also discuss the  duality cascade and chiral symmetry breaking 
in the presence of the conifold. In section 5, 
we show that quantum moduli space of the corresponding gauge theory is a 
deformed conifold. We conclude in section 6.

\section{Conifold and its type IIA brane construction: a review}
A conifold is a complex
submanifold in $\bf C^4$ described by the quadratic equation:
$       \sum^4_{i=1} z_i^2 = 0. $
Its metric is known   \cite{Candelas:1990js} to be  $ds^2= dr^2 + r^2 
ds^2_{T^{1,1}}$ with 
\beq
ds^2_{T^{1,1}} ={1\over 9}
(d\psi + \cos \theta_1 d\phi_1 + \cos \theta_2 d\phi_2)^2
+ {1\over 6} \sum^2_{i=1}
(d\theta_i^2 +\sin^2 \theta_i d\phi_i^2).
\eeq
The base of the conifold, $T^{1,1}$, is an $S^1$ bundle over
$S^2\times S^2$, and has the metric $ds^2_{T^{1,1}}$ given above.
The conifold is a Calabi-Yau manifold and respects ${\cal N}=2$ supersymmetry. 
By putting $N$ D3 branes at the tip of the 
conifold we can get ${\cal N}=1$ supersymmetric field theory living on the D3.   
This field theory was constructed by Klebanov and
Witten   \cite{KW}. It is a $SU(N)\times SU(N)$ gauge theory
\footnote{The reason for the gauge group being a product group
 is due to the underlying orbifold symmetry $Z_2$.} coupled to
two chiral superfields $A_i,\; i=1,2$ in the $(N,\bar{N})$ and   $B_j$ in the $(\bar{N}, N)$
representation. The most general superpotential which preserves
the symmetry of the conifold is \beq
  W= \Tr \epsilon^{ij} \epsilon^{kl} A_i B_k A_j B_l.
\eeq One can wrap various D branes over the cycles of $T^{1,1}$ and
identify these with states in the field theory
  \cite{Gubser:1998fp}.

A very intuitive way to understand the field theory content is to
T-dualize the theory and consider brane configurations in type IIA
string theory  \cite{Karch:1998yv,DasM,DasMtwo,Ur}.  As we discussed before, there are 
ambiguity which T-duality should we take. We focus the regular D3 branes and 
take ${\rm T}_6$ duality.
In Ref. \cite{DasM}, $N$ D3 branes on a conifold is shown to be $T_6$-dual 
to the type IIA brane configuration with 
NS5(1,2,3,4,5), NS$5'$(1,2,3,7,8,9) and  $N$ D4(1,2,3,6).  $x^6=\psi$ is periodic, NS brane is at $\psi=0$, NS$'$ is at 
 $\psi=2\pi$, $N$ D4 branes wrap the $x^6$ circle 
 so that the system is an elliptic model.
In type IIA picture it is very easy to see the resulting theory 
is $SU(N)\times SU(N)$ gauge
theory with $2N$ flavors $(A_i,B_i), i=1,2$. The theory is conformally invariant, as
can be checked explicitly from the calculation of beta functions \cite{KW}.

Introducing $M$ fractional D3 branes into the picture,  
the resulting gauge theory is $SU(N+M)\times SU(N)$ (with gauge
coupings $g_1$ and $g_2$ respectively)   \cite{kl+st,DasMtwo}. 
In type IIA picture, this corresponds to
 putting $M$  D4 branes between $0\leq \psi\leq
2\pi$ and the theory is no longer
conformally invariant. 
The two gauge couplings are
determined as follows   \cite{law,Morrison:1999cs}:
\be {1\over g_1^2} + {1\over g_2^2} \sim e^{-\phi} , \qquad 
 {1\over g_1^2} - {1\over g_2^2} \sim
e^{-\phi}\left [ \left(\int_{S^2} B_2\right) - 1/2 \right ] \ .
\ee  
Since RG flow of the coupling constant in
${\cal N}=1$ gauge theory is logarithmic, the gravity dual is expected to have 
similar logarithmic behavior in the radial coordinate of $AdS_5$
space and in fact this is true   \cite{KN}. 
In terms of the type IIA brane construction, the two gauge couplings
are determined by the positions of the NS5 branes along the $x^6$ circle.
If one of the NS5 branes is located at $x_6=0$ and the other at
$x_6=a$, then
\be \label{IIAcplgs}
{1\over g_1^2} ={l_6 - a\over g_s}\ ,\qquad
{1\over g_2^2} ={a\over g_s}\ , \label{couplings}
\ee
where $l_6$ is the circumference of the $x^6$ circle    \cite{DasM}.

As the NS5 branes approach each other, one of the couplings becomes strong.
In fact, the two gauge couplings $1/g_1^2$ and $1/g_2^2$ flow in
the opposite directions and there is a scale where one of the
couplings diverge, which necessitates the use of Seiberg's dual
gauge theory  \cite{Seiberg:1995pq}. In the present situation, this 
corresponds to the moving the NS brane  across the NS$'$ brane.
After all the  re-connections are made, we get  $SU(N-M)\times
SU(N)$ theory   \cite{kl+st}. Notice that $SU(N_f-N_c)=SU(N-M)$. 

As we go to the further IR region, the same process repeat until 
we get $SU(M+p)\times SU(p)$ with $p$ less than $M$. This is so
called the cascade of the duality. The simplest case is $p=1$. 
one may consider this as one D3 brane probing the  the background of 
M-fractional branes. Using the  ADS superpotential  \cite{Affleck:1985xz}, 
together with the original
classical superpotetial for the conifold, it was shown that 
the conifold singularity is resolved into `deformed conifold'  \cite{kl+st}.
A very interesting phenomenon happens as a consequence, namely 
the chiral symmetry is broken.

\section{Orientifold in conifold}
The orientifolding operation is consist of world sheet orientation 
reversal and reflection of transverse spacetime, together with 
appropriate projection of the left moving fermion number. 
We will concentrate on the spacetime reflection part hereon.  
In the conifold there are  many $Z_2$ operations  \cite{KW} so that it is not 
clear which is the relevant $Z_2$ for the orientifolding.  
Therefore 
we want to derive the orientifold operation in the conifold from 
the type IIA picture where O4-plane is clearly defined.

We have D4 (01236), O4 (01236), NS5 (012345), NS5'(012389) with  $x^6$ compact.
Under the T-duality along the $x^6$, the whole
configuration is  mapped to the conifold.  
The prescription for the (T-duality) mapping flat $R^{6}$ 
to  conifold  suggested by Dasgupta and  Mukhi  \cite{DasM}  is: 
\ba 
 x^4,x^5 \; 
{\rm plane}  &\to& S^2 \;\; {\rm described ~~~ by}\;\;
\theta_1,\phi_1, 0 \leq \theta_i <\pi, 0\leq \phi_i < 2\pi, \no
x^8,x^9 \; {\rm plane}  &\to& S^2 \;\; {\rm described ~~~ by}\;\;
\theta_2,\phi_2, \no
 x^6 &\to& \psi  , \; \;\;\; |x^7| \to \log 1/r.
\ea

The orientifolding in type IIA picture is given by
\be
x^i\to -x^i \;\;{\rm for}\; i=4,5,7,8,9, \ {\rm and} \; x^i\to x^i,\  {\rm otherwise}.
\label{ofd}
\ee 
Under T-duality  along $x^6$,  O4 brane becomes O3 brane. 
The reflections in $x^4,x^5,x^6,x^8,x^9$ induces an antipodal
mapping in $S^2$ and $S^3$ of $T^{1,1}$, the base of the cone.
Above reflection in terms of the polar coordinates is
\be
\phi_i\to\phi_i+\pi_i, \;\;  \theta_i \to \pi-\theta_i, \;\; \psi\to -\psi.
\ee

Now we  want to  express the above operation in terms of the 
conifold variables  $z_i$'s. For doing that we have to express them in terms of the angular variables. 
Fortunately this has been done in Ref.\cite{Candelas:1990js}.  
Introducing the variable
\be
Z\equiv \pmatrix{
  z_{11} & z_{12} \cr
  z_{21} & z_{22} }
  \equiv \frac{1}{\sqrt{2}} \pmatrix{
  z_3+iz_4  & z_1-iz_2 \cr
  z_1+iz_2 & -z_3+iz_4 }
  , \ee
the equation defining the conifold  can be rewritten as
$\det (z_{ij})=0$.
The base $T^{1,1}$ is the intersection of
the conifold with the $S^5$ sphere $ \sum_{i=1}^4 |z_i|^2=r^2$,
and is locally $SU(2)\times SU(2)/U(1)$. 
We  parametrize $T^{1,1}$ by $SU(2)$ parameters $a_i,b_i, \; i=1,2$ :
\be
\pmatrix{
  a_i   \cr
  b_i }  = \pmatrix{
  \cos\frac{\theta_i}{2} e^{i(\psi+\phi_i)/2} \cr
   \sin\frac{\theta_i}{2} e^{i(\psi-\phi_i)/2} }.
\ee
We can naturally map the $S^2$'s, which are compactifications  of the 
4,5 and 8,9 planes, into two $SU(2)$'s by the Hopf fibration 
$ {\vec n} = {\bf a}^\dagger {\vec \sigma} {\bf a} $,
where ${\vec n} \in S^2$, ${\bf a}= \pmatrix{  a   \cr   b }$.
With this identification, the antipodal mappings in the two $S^2$ 
induces a mapping in $T^{1,1}$ hence a mapping in the conifold.
From the two $SU(2)$ matrices 
$L= \pmatrix{
  a_{1} & -{\bar b}_{1} \cr
  b_{1} & {\bar a}_{1} } ,R= \pmatrix{
  a_{2} & -{\bar b}_{2} \cr
  b_{2} & {\bar a}_{2} }$,  we may construct the base
variable
\be
Z/r=LZ_0R^\dagger =
 \pmatrix{
    -a_1 b_2  & a_1a_2 \cr
    -b_1 b_2 & b_1a_2 },
\ee
with $r^2=\Tr(ZZ^\dagger) =\sum_i \abs{z_i}^2$.

In terms of $SU(2)$ variables the orientifold operation is 
\be a_i
\to i{\bar b}_i,\; b_i\to -i{\bar a}_i, \label{oinb}\ee
 which implies $ z_{11} \to -{\bar
z}_{22}, \; \; z_{12} \to {\bar z}_{21}$.  In terms of original
variables $z_i$'s, above operation has a simple expression: 
\be z_i \to {\bar z}_i , \; i=1,\cdots,4. \label{oincon} 
\ee 
The conifold is invariant under this $Z_2$ operation and so is the
superpotential.

Let $z_j=x_j+iy_j$, $j=1,\cdots, 4$ with $x_j,y_j$ be real. 
Then  fixed points of this
reflection are given  by $y_i=0, \; i=1,\cdots,4$. Together with
the conifold equation $\sum_i z_i^2=0$, we conclude that $z_i=0$,
i.e. the tip of the cone, is the only fixed point. This is
consistent with the fact that the O3 is a point in the conifold.

However, O3 branes are  like fractional D3 branes, 
since they are between NS branes rather than wrapping the 
whole circle of the IIA picture. 
In type IIB conifold picture, they are O5 branes wrapping the 
different singular $S^2$ cycles. 
In fact the base of the $T^{1,1}$ is $\frac{SU(2)\times 
SU(2)}{U(1)}$ and the $U(1)$ is acting symmetrically on both 
$SU(2)$'s \cite{Candelas:1990js} by 
$$ \pmatrix{
    e^{i\theta}  & 0 \cr
    0 & e^{-i\theta}  }.
$$
It is also known that the 
second homology basis is a combination of  two $S^2$'s above, namely,  $\Sigma_2 =S^2_2 -S^2_1$ \cite{Gubser:1998fp,DasMtwo}. 
The charge of the O3 brane is determined according to which $S^2$ 
the O5 wraps.
Therefore both $O3^+$ and $O3^-$ can be considered as a wrapped $O5$ 
branes wrapping different vanishing cycles. 
Since they are wrapping different $S^2$'s (although in the vanishing limit),
we can say that they are stable due to the topological reason. 
However, since they are wrapping vanishing cycles at the same 'point', they are like 
overlapping charges of opposite charge.
So if we measure the net effect, there are no untwisted RR charges and it is  an effect  
of one object that has only twisted RR charge. This situation is closely related to the 
orientifold of $C^2/Z_N$ considered by Uranga \cite{uranga3}

We now ask  what happen to the small resolution of the conifold?
\be 
\abs{a_1}^2 +\abs{b_1}^2- \abs{a_2}^2 -\abs{b_2}^2 =\delta .
\ee
One immediately see that the small resolution of the conifold is 
invariant under the orientifold operation.  However, from  Eq.(\ref{oinb})
 there is no fixed point on the $S^2$ 
unless its size is zero. Only when  the fractional branes are  
wrapping the 'vanishing cycle', we can have orientifold in 
conifold. This means that the resolved conifold does not admit an 
super symmetric orientifold. 

If there are large number of fractional branes, 
due to the quantum effect, the conifold background is modified \cite{kl+st} to
deformed conifold $\sum_i z_i^2 = \mu$. In this case,
the fixed points form a manifold $S^3$ given by $\sum_i x_i^2 
=\mu$. Does this mean that O6 is created? 
Then, where is the orientifold in the deformed conifold? 
Part of the answer also lies in the large $N$ geometric transition   
\cite{Maldacena,vafa}: when the 
number of the fractional branes are large, it goes to the large N dual 
description where branes disappear and only flux remains. 
Presumably O3 branes disappeared leaving only its flux. 
Since O3$^-$ comes with extra two  D3, $O3^\pm$ contribute the same amount of 3-form flux.   
Therefore the equality $O3^+=O3^-+ 2D3$ holds  as far as supergravity solutions are 
concerned. The flux of the fractional D3 brane charge resolve the 
singularity in the $O3^+$ side.  Since O6 RR charge can not be created 
by the deformation process, the most natural answer to above question seems to be that the $O3^+$ 
charge is smeared uniformly over the fixed manifold, while $O3^-$
is wrapping the vanishing $S^2$. We will be back to this issue in 
the discussion section  for other possibilities. 
Now we make some remarks. 
\begin{enumerate}
\item
In type IIA, $x^7$ is a spectator. However, under the
identification $|x^7| =\log 1/r$, we are abandoning the region $r\geq 1$. 
We could equally cut out the $r\leq 1$ by the identification $|x^7| =\log r$.
We have chosen above convention since
we are interested in the near the singularity:  $r \to 0$
corresponds to $x^7 \to \infty$.  

\item 
Recently the same operation was
considered in   Ref.\cite{Sinha:2000ap} in the context of the
topological string with  O3 brane wrapping $S^3$. Here
our interpretation is different: The fixed three sphere
of the deformed conifold is due to smearing of the O3 brane charge.
\item It may be worth while to describe the orientifolding induced by the $T_\phi$ duality,
 although we do not follow it. In this scheme, 4,5 and 8,9 plane 
 is consequence of the T-duality along the circle action at the two sphere of the resolved conifold: 
 two fixed points of the circle action under the $T_\phi$ duality is 
 mapped to the NS and NS$'$ branes. The orientifold operation of the 
 type IIA branes does not involve the reflection along $x^6=\psi$ 
 therefore not a antipodal point mapping. The action is simply 
 given by 
 \be \phi_i\to \phi_i +\pi, \;\; \theta_i \to \theta_i, \psi \to \psi. \ee 
 In terms of $a_i$ and $b_i$, this is written 
 \be
 a_i \to a_i ,\;\; b_i \to -i b_i.
 \ee
 Finally, in terms of the $z_i$'s, 
 \be z_1 \to z_1, \;  z_2 \to z_2, \;  z_3 \to -z_3, \; z_4 \to -z_4. 
 \ee
 which also appear in literature as an orientifold operation.
\end{enumerate}

\section{The RG analysis and Duality cascade in the presence of the orientifold}
If $N$ D3 branes are sitting at the singular point of the
conifold, the gauge theory content turned out to be $SU(N)
\times SU(N)$   \cite{KW}. In this section,  we discuss  how introducing the
orientifold modify the theory. 

We use the
type IIA brane picture.
If we add O4 plane  along 01236 direction,
we have to change the sign of the RR charge of the O4 plane as we
cross the NS brane  \cite{Evans:1997hk,Giveon:1999sr}. The corresponding gauge theory must 
be of 
alternating SO and Sp gauge groups. So one may naively
expect that the gauge group would be  $SO(N)\times Sp(N)$.
(Here we use the convention where $Sp(2)=SU(2)$.)  However, this
is not conformal as we can see from the RG flow:
 \ba
{d\over d\log(\Lambda/\mu)}  \left[{8\pi^2 \over g_{so}^2}\right] &\sim & 
3(N-2)-2N(1-\gamma),
\no
{d\over d\log(\Lambda/\mu)} \left[ {8\pi^2 \over g_{sp}^2}\right] &\sim & 3(N/2+1)-(N)(1-\gamma)
.
\ea

One cannot require  two beta functions to vanish
simultaneously. We can overcome the difficulty if we
replace $SO(N)$ by $SO(N+2)$. In brane language, we
should add an extra brane and its image over the negative charged
orientifold for the stable configuration.

The presence of the extra two branes can be also  understood from the brane 
dynamics \cite{Evans:1997hk}: 
for stable configuration, RR charge density must be continuous across the NS5.
Since O4$^\pm$ has RR charge $\pm 1$, we need extra 2 D4 over the O4$^-$.
Then the gauge group is $SO(N+2)\times Sp(N)$.
 One may interpret the extra 2 branes as the fractional branes in
IIB picture \cite{DasMtwo,kl+st}. So in the presence of the orientifold,
 the  brane configuration with no fractional branes is not stable fixed point.

If we place $N$ D3 branes and $M+2$ fractional D3 branes on
the conifold with an O3 plane,
we obtain an $SO(N+M+2)\times Sp(N)$ or $Sp(N+M)\times SO(N+2)$ gauge groups.
Here, we discuss the first one since the other one is 
exactly parallel. The two gauge group factors have
holomorphic scales $\Lambda_1$ and $\widetilde{\Lambda_1}$. 
The matter consists of two chiral superfields $A_1,A_2$ in the
$({\bf N+M+2,\overline{ N} })$ representation and two fields $B_1,B_2$ in the
$({\bf {N+M+2},N})$ representation.  
The invariants of the theory under the global symmetry $SU(2)\times SU(2)\times U(1)$ are  
\begin{equation}
I_1\sim \lambda_1^{3M+2}{\widetilde{\Lambda_1}^{2b_{Sp}} \over \Lambda_1^{b_{SO}}}
\ [\tr\ (A_iB_jA_kB_\ell\epsilon^{ik}\epsilon^{j\ell})]^{2M},
\end{equation}
\begin{equation}
R_1^{(1)}  = {\tr\ [A_iB_j]\tr[A_kB_\ell]\epsilon^{ik}\epsilon^{j\ell}
\over \tr\ (A_iB_jA_kB_\ell\epsilon^{ik}\epsilon^{j\ell})} \ ;
\end{equation}
\begin{equation}
J_1\equiv \lambda_1^{b_{SO}+2b_{Sp} }\Lambda_1^{b_{SO}}
\widetilde{\Lambda_1}^{2b_{Sp}},
\end{equation}
where ${b_{SO}}$, ${b_{Sp}}$ are $\beta$ function coefficients:
\ba
{b_{SO}}&=& 3(N+M+2 -2)- 4\cdot \frac{N}{2} \cdot 1 =N+3M, \no
{b_{Sp}}&=& 3(\frac{N}{2}+1)- 4\cdot \frac{N+M+2}{2} \cdot \half 
=N/2-M+1.
\ea  
These statement can easily be verified by the quantum number 
assignment similar to that given in table 1 in Ref.\cite{kl+st}. The only difference 
 is that $\Lambda^{3(N+M)-2N}$ and ${\tilde\Lambda}^{3(N)-2(N+M)}$
 of the for $SU(N)\times SU(N)$ should be replaced by $\Lambda^{3(N+M+2-2)-2N}$ 
 and ${\tilde\Lambda}^{3(N+2)-2(N+M+2)}$ of the $SO(N+M+2)\times  Sp(N)$.
The superpotential of the model will get multiplicative renormalization 
depending on $I_1,J_1,R_1$'s. 

In the presence of the orientifold, 
the geometry of the base of the cone is $RP^2 \times S^3$.
The 3-form flux breaks the conformal symmetry and $B_2$ gets radial dependence.
\be
\int_{{\rm\bf RP}^2} B_2 \sim (M/2)  \ln (r/r_0),
\ee
which indicate the logarithmic running of ${1\over g_{SO}^2} - {1\over
g_{Sp}^2}$ in the $SO(N+M+2)\times Sp(N)$ gauge theory.
To check this bulk result, we look at the $ \beta$ functions of the boundary theory:
 \ba
{d\over d\log(\Lambda/\mu)}  \left[{8\pi^2 \over g_{so}^2}\right] &\sim & 3(N+M)-2N(1-\gamma)
\no
{d\over d\log(\Lambda/\mu)} \left[ {8\pi^2 \over g_{sp}^2}\right] &\sim & 3(N/2+1)-(N+M+2)(1-\gamma)
\ea
For the conformal invariance of $M=0$ case, we require $\gamma = -1/2 $.
Notice that two flows give
 the same condition of the anomalous dimension in spite of the
 difference of the gauge group.
The difference of the flow is
\be
{8\pi^2 \over g_{SO}^2} -{8\pi^2 \over g_{Sp}^2} = (4-\gamma)M 
\log(\Lambda/\mu).
\ee
showing the the consistency of bulk and boundary result.

Since 
$1/g_{SO}^2$ and $1/g_{Sp}^2$ flow in opposite directions, 
there is a scale where the
$SO(N+M+2)$ coupling, $g_{SO}$, diverges. Using the 
 Seiberg duality transformation, we get the $SO(2N-[N+M+2]+4) = SO(N-M+2)$ 
gauge group with $2N$ flavors $(a_i,b_i)$ and ``meson'' bilinears
$M_{ij}= A_iB_j$.  The fields $a_i$ and $b_i$ are anti-fundamentals
and fundamentals of $Sp(N)$, while the mesons are in the
adjoint-plus-singlet of $Sp(N)$.  The superpotential after the
transformation
\begin{equation}
W = \lambda_1 \tr\ M_{ij}M_{k\ell}\eps^{ik}\eps^{j\ell}F_1(I_1,J_1,R_1^{(s)})
+ {1\over \mu} \tr\  M_{ij} a_ib_j \ ,
\end{equation}
where $\mu$ is the matching scale for the duality transformation
  \cite{kl+st}, shows the $M_{ij}$ are actually massive. Thus we may 
integrate them out and get the superpotential
\begin{equation}
W = \lambda_2
\tr\ a_i b_j a_k b_\ell \eps^{ik}\eps^{j\ell} F_2(I_2,J_2,R_2^{(s)}).  
\end{equation}
Here $F_2$, $\lambda_2$, $ I_2$, $J_2$ and $R_2$ are defined
similarly as in the original theory.  Thus we obtain an $
Sp(N) \times SO(N-M+2)  $
theory which resembles closely the theory we started with.
This can be shown more carefully using the matching condition as 
discussed in   \cite{kl+st}. 
The next step is that the $Sp(N)$ gauge group
becomes strongly coupled, and under a Seiberg duality transformation
the full gauge group becomes $SO(N-3M+2)\times Sp(N-4M)$, and so forth.

\section{Deformation of the orientifolded conifold}
The  classical field theory reveals that it represents branes
moving on a conifold   \cite{KW,Morrison:1999cs} by having the conifold as its moduli space of the 
gauge theory. 
Klebanov and Strassler showed that the vacuum configurations of 
gauge theory, 
the conifold, is modified to a deformed conifold by studying 
non-perturbative quantum corrections   \cite{kl+st}.
Here we study corresponding phenomena in the presence of the orientifold. 
In our case, the minimal configuration is  $SO(M+4)\times Sp(2)$, corresponding
to a D-brane and its image in the presence of $M+2$ fractional branes.
The fields  are
\ba
A_{i,\alpha}^r,  B_{j, \alpha}^a: i=1,2, \quad \alpha=1, ..., N_c,
\quad r=1, ..., N_f.
\ea
Define $N_{ij}$ by $ N_{ij}^{rs} =\sum_{\alpha} A_{i, \alpha}^r B_{j, 
\alpha}^s$.
Then the classical superpotential can be written as
\begin{equation}
W_c = \half \lambda \mbox{Tr} A_i B_j A_k B_l \epsilon^{ik} \epsilon^{jl} =
\half \lambda \mbox{Tr} N_{ij} N_{kl}  \epsilon^{ik} \epsilon^{jl} .
\end{equation}
where the trace is over the flavor index. 
For the gauge theory with $N_c$ colors and $N_f$ flavors, the
quantum effect gives the ADS super potential   \cite{Affleck:1985xz}.
In our case, the gauge group is $SO(N+M+2) \times Sp(N)$. So one
may expect that we should add up the ADS potential for each gauge
group. However, one should notice that we are interested in
$ M \gg 2=N$ and the ADS potential is meaningful only for $N_c > 
N_f$. So we only have to consider the superpotential for the $SO(N+M+2)$ part.
We propose that following is the leading order superpotential for the
product group:
\begin{equation}
W_{total} = \lambda  W_{c}  +
({N_c-2-N_f}) \left( \frac{\Lambda^{b_0}}
{\det _{ir;js}N_{ij}^{rs} } \right)^{1/(N_c-2-N_f)}.
\end{equation}

Now we consider the completely  higgsed configuration
$SU(N)\to U(1)^N$,
 where the matrix $N_{ij}$ have vacuum expectation values:
\begin{equation}
N_{ij}^{rs} =  n_{ij}^{(r)}\delta^{rs}.
\end{equation}
This is justified since $r$-th and $s$-th fractional (flavor) branes are 
far separated in higgsed case. 
So, the determinant is trivially factorized:
\begin{equation}
\mbox{det}_{ir,js}N_{ij}^{(rs)}=  \prod_{r=1}^{N_f} W^{(r)},
\end{equation}
where $ W^{(r)}= n^{(r)}_{ij} n^{(r)}_{kl}  \epsilon^{ik} \epsilon^{jl}$
Using $W_c =\lambda \sum_r W^{(r)},$
and considering $N_f=2$  case, the total superpotential becomes
\begin{equation}
W_{total} = \lambda ( W^{(1)} + W^{(2)} ) +
 (N_c-2-N_f) \left( \frac{\Lambda^{b_0}}
{W^{(1)} W^{(2)} } \right)^{1/(N_c-2-N_f)},
\end{equation}
where $b_0=3(N_c-2)-N_f$.
The total potential is minimized if we have 
\begin{equation}
(N_{ij}N_{kl}\epsilon^{ik}\epsilon^{jl})^{(N_c-N_f)} =
\left( \frac{\Lambda^{b_0}} {\lambda^{Nc-N_f-2} }\right).
\end{equation}
Notice that $N_c-N_f=(M+2)-2=M$ in our case.
This is nothing but the equation for the deformed conifold and there are $M$ branches:
each of the probe branes move on the deformed conifolds.

The M-theory curve  \cite{Witten:1997ep} for the Type IIA version is given as follows:
\be
(vw)^{M}=\Lambda^{3M}, \quad t=v^{N_c-N_f}=v^M .
\ee
The curve indicates that if there is more than two fractional
branes ($M>0$), the whole brane configuration is nicely connected
and this indicates the deformation of the conifold.
Interesting fact is that the curve is insensitive to the presence
of the regular D branes  since it contribute to $N_f$ as well as to $N_c$.
It depends only on the number of fractional branes.  Orientifolds 
does not change the formal behavior at all, either. It just require that $M$ is even. 

 \section{Conclusion}
In this paper, we discussed the  string theory in  conifold in the presence of the
orientifold.   We first mapped the orientifolding operation from the type IIA brane picture 
to the conifold picture, using the T-duality along the $x^6$ direction.
 Under the T-duality  
 the fractional D4 branes ( D4 branes between NS-NS$'$ branes) are mapped to
  fractional D3 branes, which is identified as   D5 wrapping the vanishing 
  2-cycles. 
We showed how the conifold can admit orientifolds of both $+$ and $-$ 
charges. Blowing up the tip, the base is still $S^2\times S^2 \times S^1$.  The 
fractional $D3$  and $O3^-$ are D5 and $O5$ branes  respectively  wrapping one of the $S^2$.
The  $O3^+$ is O5 brane wrapping the other $S^2$. Since  
homology 2-cycle which defines the RR charge  is given by the difference of two $S^2$'s, it is 
natural to have both  $O3^+$ and  $O3^-$ simultaneously at the tip of the cone without annihilating each
other: they are topologically protected not to be annihilated. 
We then showed that in the presence of the orientifold, the conformally invariant configuration
is $SO(N+2)\times Sp(N)$ rather than $SO(N)\times Sp(N)$. This
is shown both by field theory and brane dynamics. If we add
fractional branes, there are duality cascade as in  Ref.\cite{kl+st}. 
We analyzed the corresponding gauge theory as well as the super gravity 
solution. When there are fractional branes, the conifold is deformed as is
the case of $SU(N)\times SU(N)$. We showed this by writing quantum
corrected superpotential for the product group.

However more detailed discussion of the supergravity part is not discussed. 
We expect that the gravity solution is the same as Klebanov and Tsyetlin since 
it cannot distinguish the orientifolds and  D branes.   
However we expect that there are minor modification due to torsion 
part.  Also the K-theoretic discussion of the 
orientifold charge in the conifold is also worthwhile do be discussed in detail. 
We wish to come back to this issue in near future. 

Finally we give a discussion on the deformation of the orientifold.
As we have seen in section 6, the gauge theory analysis showed that 
the conifold is deformed by the quantum mechanical effect.
First we should notice that, once the conifold is deformed, 
the connection between $z_{ij}$ and $(a_i,b_i)$ is lost.
So there is no way to derive the orientifold operation in the deformed  conifold 
from that of the IIA branes. 
What we have done is to {\it assume} that the orientifold operation of the original (singular) 
conifold  $z_i \to {\bar z_i} $ is still valid for the deformed case.
Then $S^3$ is the fixed points, which gave us a big conceptual 
trouble, since the most natural interpretation is to regard $S^3$ as the cycle wrapped by O6. 
But this is not allowed by the SUSY of IIB. 
In this situation, we have several options. 
\begin{enumerate}
\item The identification of orientifold operation  $z_i \to {\bar z_i} $, for the deformed case,
 is not correct.
\item  O3 charge is smeared out 
so that it has O3 charges but have the effect of the O6. In this case, the SUSY is intact.
\item  O6 really is created by the deformation.  In this case we loose SUSY. 
\end{enumerate}
We excluded the first option since  $z_i \to {\bar z_i} $  
is the only one that leads to the correct one in the limit of zero deformation parameter.  
We have chosen the second option since 
gauge theory or SUGRA so far does not show any signal of supersymmetry breaking.

However, before the paper of Strassler and Klebanov \cite{kl+st} appeared, Mukhi and 
 DasGupta in  \cite{DasMtwo} claimed that SUSY is broken when the singularity is resolved. 
 So the situation not entirely clear even in the absence of the orientifold
  and we believe that it is an interesting subject to study. 
One may further speculate that the appearance of fixed $S^3$ means appearance of O6 
meaning the supersymmetry is broken. If true it can be used as a dynamical supersymmetry  breaking 
mechanism.  However,  this require more extensive analysis which goes beyond the 
scope of present work and we hope to settle this issue in later 
publication.

\noindent{\bf \large Acknowledgements}  \\ \\
We would like to thank I. Klebanov and Uranga for discussion and APCTP for support and 
hospitality during our visit. 
The work of SJS is supported by KOSEF 1999-2-112-003-5. The work of
CA and SN is supported by KOSEF 2000-1-11200-001-3.
This work is also supported by BK21 program of Korea Research Fund (2001).


\newpage
\begin{thebibliography}{999}

\bibitem{Maldacena} J. Maldacena, Adv. Theor. Math. Phys. {\bf 2}
(1998) 231 [hep-th/9711200].

\bibitem{GKP} S.S. Gubser, I.R. Klebanov, and A.M. Polyakov,
Phys. Lett. {\bf B428} (1998) 105 [hep-th/9802109].

\bibitem{WittenADS} E. Witten, Adv. Theor. Math. Phys. {\bf 2} (1998) 253  [hep-th/9802150].

\bibitem{Candelas:1990js}
P.~Candelas and X.~C.~de la Ossa, 
Nucl.\ Phys.\  {\bf B342} (1990) 246. 

\bibitem{Kehagias} A. Kehagias, Phys. Lett. B435 (1998) 337 [hep-th/9805131].

\bibitem{KW}  I.R. Klebanov and E. Witten, Nucl. Phys. {\bf B536} (1998) 199 [hep-th/9807080].

\bibitem{kl+st} I.R. Klebanov and M.J. Strassler, JHEP {\bf 0008} (2000) 052 [hep-th/0007191].


\bibitem{Sagnotti}
G.~Pradisi and A.~Sagnotti,
Phys.\ Lett. {\bf B216} (1989) 59; 
M.~Bianchi and A.~Sagnotti,
Phys.\ Lett. {\bf B247} (1990) 517;
M.~Bianchi and A.~Sagnotti,
Nucl.\ Phys. {\bf B361} (1991) 519.


\bibitem{Gimon:1996rq}
E.~G.~Gimon and J.~Polchinski,
Phys.\ Rev.\  {\bf D54} (1996) 1667 [hep-th/9601038].
 
\bibitem{Park:2000ep}
J.~Park, R.~Rabadan, and A.~M.~Uranga,
Nucl.\ Phys.\ {\bf B570} (2000) 38
[hep-th/9907086].

\bibitem{oh}K.~Oh and R.~Tatar,
JHEP {\bf 0005} (2000) 030 [hep-th/0003183]. 

\bibitem{ohhyun} K.~Dasgupta, S.~Hyun, K.~Oh, and R.~Tatar,
JHEP {\bf 0009} (2000) 043 [hep-th/0008091]. 

\bibitem{Sinha:2000ap} S.~Sinha and C.~Vafa,
``$SO$ and $Sp$ Chern-Simons at large $N$,''
hep-th/0012136. 

 
\bibitem{Naculich:2001xu}
S.~G.~Naculich, H.~J.~Schnitzer, and N.~Wyllard,
``$1/N$ corrections to anomalies and the $AdS$/CFT correspondence for  orientifolded $N = 2$ orbifold models and $N = 1$ conifold models,''
hep-th/0106020.


\bibitem{DasM} K. Dasgupta and S. Mukhi,
 Nucl. Phys.  {\bf B551} (1999) 204 [hep-th/9811139].

\bibitem{Gubser:1998fp} S.~S.~Gubser and I.~R.~Klebanov,
Phys.\ Rev.\  {\bf D58} (1998) 125025 [hep-th/9808075].

\bibitem{uranga3} Angel M. Uranga, 
 Nucl.Phys. B577 (2000) 73-87 [hep-th/9910155].

\bibitem{Karch:1998yv}
A.~Karch, D.~Lust and D.~Smith,
Nucl.\ Phys.\ {\bf B533}, 348 (1998)
[hep-th/9803232].


\bibitem{DasMtwo} K.~Dasgupta and S.~Mukhi,
JHEP {\bf 9907} (1999) 008 [hep-th/9904131]. 


\bibitem{Ur} A.M. Uranga, JHEP {\bf 9901} (1999) 022 [hep-th/9811004].

\bibitem{Ooguri:1996wj} H.~Ooguri and C.~Vafa,
Nucl. \ Phys.\  {\bf B463} (1996) 55 [hep-th/9511164].


\bibitem{law}
A.~E.~Lawrence, N.~Nekrasov and C.~Vafa,
Nucl.\ Phys.\  {\bf B533} (1998) 199 [hep-th/9803015].

\bibitem{Morrison:1999cs}
D.~R.~Morrison and M.~R.~Plesser,
Adv.\ Theor.\ Math.\ Phys.\ {\bf 3} (1999) 1 [hep-th/9810201].

\bibitem{KN} I.R. Klebanov and N.A. Nekrasov, Nucl. Phys.  {\bf B574} (2000) 263 [hep-th/9911096].


\bibitem{Seiberg:1995pq} N.~Seiberg,
Nucl.\ Phys.\  {\bf B435} (1995) 129 [hep-th/9411149].

\bibitem{Affleck:1985xz} I.~Affleck, M.~Dine, and N.~Seiberg, 
Nucl.\ Phys.\  {\bf B256} (1985) 557. 


 
\bibitem{vafa}  C. Vafa,   
``Superstrings and Topological Strings at Large $N$,"  hep-th/0008142,
 ;  M. Atiyah, J. Maldacena, C. Vafa, 
  ``An M-theory Flop as a Large $N$ Duality," hep-th/0011256. 
 
\bibitem{Witten:1997ep}
E.~Witten,
Nucl.\ Phys.\  {\bf B507} (1997) 658 [hep-th/9706109].
 

\bibitem{Giveon:1999sr} A.~Giveon and D.~Kutasov, 
Rev.\ Mod.\ Phys.\ {\bf 71} (1999) 983 [hep-th/9802067].


\bibitem{Evans:1997hk}
N.~Evans, C.~V.~Johnson and A.~D.~Shapere,
Nucl.\ Phys. {\bf B505}  (1997) 251
[hep-th/9703210].
 
  

\end{thebibliography}
\end{document}